\documentclass[aps,prd,onecolumn,groupedaddress,showpacs,nofootinbib,amssymb]{revtex4-2}
\usepackage[dvips]{graphicx}
\usepackage{amssymb}
\usepackage{amsmath}
\usepackage{graphicx,,color}
\usepackage{amsfonts}
\usepackage{bm}
\usepackage{cancel}
\usepackage{comment}

\newcommand\be{\begin{equation}}
\newcommand\ee{\end{equation}}

\allowdisplaybreaks[4]

\begin{document}

\tolerance=5000

\title{Spectrum of Primordial Gravitational Waves in Modified Gravities: A Short Overview}
\author{S.D. Odintsov,$^{1,2,3,6}$\,\thanks{odintsov@ieec.uab.es}
V.K. Oikonomou,$^{4}$\,\thanks{v.k.oikonomou1979@gmail.com},R.
Myrzakulov$^{5,6,7}$}
\affiliation{$^{1)}$ ICREA, Passeig Luis Companys, 23, 08010 Barcelona, Spain\\
$^{2)}$ Institute of Space Sciences (ICE,CSIC) C. Can Magrans s/n,
08193 Barcelona, Spain\\
$^{3)}$ Institute of Space Sciences of Catalonia (IEEC),
Barcelona, Spain\\
$^{4)}$Department of Physics, Aristotle University of Thessaloniki, Thessaloniki 54124, Greece\\
$^{5}$ Eurasian National University, Nur-Sultan 010008, Kazakhstan\\
$^{6}$ Ratbay Myrzakulov Eurasian International Centre for Theoretical Physics, Nur-Sultan 010009, Kazakhstan\\
$^{7}$ Laboratory for Theoretical Cosmology, International
Centre of Gravity and Cosmos, (TUSUR) , 634050 Tomsk, Russia\\
}

\begin{abstract}
In this work we shall exhaustively study the effects of modified
gravity on the energy spectrum of the primordial gravitational
waves background. S. Weinberg has also produced significant works
related to the primordial gravitational waves with the most
important one being the effects of neutrinos on primordial
gravitational waves. With this sort review, our main aim is to
gather all the necessary information for studying the effects of
modified gravity on primordial gravitational waves in a concrete
and quantitative way and in a single paper. After reviewing all
the necessary techniques for extracting the general relativistic
energy spectrum, and how to obtain in a WKB way the modified
gravity damping or amplifying factor, we concentrate on specific
forms of modified gravity of interest. The most important
parameter involved for the calculation of the effects of modified
gravity on the energy spectrum is the parameter $a_M$ which we
calculate for the cases of $f(R,\phi)$ gravity,
Chern-Simons-corrected $f(R,\phi)$ gravity,
Einstein-Gauss-Bonnet-corrected $f(R,\phi)$ gravity, and higher
derivative extended Einstein-Gauss-Bonnet-corrected $f(R,\phi)$
gravity. The exact forms of $a_M$ is presented explicitly for the
first time in the literature. With regard to
Einstein-Gauss-Bonnet-corrected $f(R,\phi)$ gravity, and higher
derivative extended Einstein-Gauss-Bonnet-corrected $f(R,\phi)$
gravity theories, we focus on the case that the gravitational wave
propagating speed is equal to that of light's in vacuum. We
provide expressions for $a_M$ expressed in terms of the cosmic
time and in terms of the redshift, which can be used directly for
the numerical calculation of the effect of modified gravity on the
primordial gravitational wave energy spectrum.
\end{abstract}

\maketitle

\section*{Introduction}

To date, the current perception of our Universe indicates that the
Universe went through four distinct evolutionary eras, the
inflationary era
\cite{inflation1,inflation2,inflation3,inflation4}, the radiation
domination, the matter domination era and the dark energy eras.
Our knowledge is limited though, since we known very well only the
physics beyond the recombination era, nearly at redshift $z\sim
1100$ where the Cosmic Microwave Background (CMB) modes exited the
Hubble horizon. Concerning the dark energy era, we do not know
what is the origin of this late-time acceleration era, and the
same applies for the inflationary era and the most mysterious of
all, the reheating and the subsequent radiation domination eras.
With regard to the inflationary era, we do not even know whether
it even occurred. Modified gravity in its various forms
\cite{reviews1,reviews2,reviews3,reviews4,reviews5,reviews6} can
play a prominent role toward consistently describing inflation and
the dark energy eras with or without the need for scalar fields.
In some cases, it is possible to describe inflation and the dark
energy eras within the same theoretical framework, see
\cite{Nojiri:2003ft} for the first attempt toward this research
line, and also Refs.
\cite{Nojiri:2007as,Nojiri:2007cq,Cognola:2007zu,Nojiri:2006gh,Appleby:2007vb,Elizalde:2010ts,Odintsov:2020nwm,Oikonomou:2020qah,Oikonomou:2020oex}
for some more recent works. However, admittedly the radiation
domination era, and specifically, its early stages, remains quite
mysterious and to date inaccessible by any currently undergoing
experiment. Hopefully though, all the future interferometric
gravitational wave experiments like the LISA laser interferometer
space antenna \cite{Baker:2019nia,Smith:2019wny}, the dHz probing
DECIGO \cite{Seto:2001qf,Kawamura:2020pcg}, the Hz-kHz frequencies
probing Einstein Telescope \cite{Hild:2010id}, and the future BBO
(Big Bang Observatory) \cite{Crowder:2005nr,Smith:2016jqs}, are
expected to probe the frequency range corresponding to the
reheating and radiation domination era. Specifically, the
frequency range of the future gravitational waves experiments
corresponds to way higher frequencies compared to the CMB ones,
and will probe modes that became subhorizon right after the
inflationary era, during the early stages of the reheating era. At
intermediate frequencies, very promising results may be obtained
by the Square Kilometer Array (SKA) \cite{Bull:2018lat} and the
NANOGrav collaboration \cite{Arzoumanian:2020vkk,Pol:2020igl},
which are based on measurements of pulsar timing arrays. In the
literature there exist several theoretical attempts to efficiently
predict the energy spectrum of the primordial gravitational waves,
see Refs.
\cite{Kamionkowski:2015yta,Denissenya:2018mqs,Turner:1993vb,Boyle:2005se,Zhang:2005nw,Schutz:2010xm,Sathyaprakash:2009xs,Caprini:2018mtu,
Arutyunov:2016kve,Kuroyanagi:2008ye,Clarke:2020bil,Kuroyanagi:2014nba,Nakayama:2009ce,Smith:2005mm,Giovannini:2008tm,
Liu:2015psa,Zhao:2013bba,Vagnozzi:2020gtf,Watanabe:2006qe,Kamionkowski:1993fg,Giare:2020vss,Kuroyanagi:2020sfw,Zhao:2006mm,
Nishizawa:2017nef,Arai:2017hxj,Bellini:2014fua,Nunes:2018zot,DAgostino:2019hvh,Mitra:2020vzq,Kuroyanagi:2011fy,Campeti:2020xwn,
Nishizawa:2014zra,Zhao:2006eb,Cheng:2021nyo,Nishizawa:2011eq,Chongchitnan:2006pe,Lasky:2015lej,Guzzetti:2016mkm,Ben-Dayan:2019gll,
Nakayama:2008wy,Capozziello:2017vdi,Capozziello:2008fn,Capozziello:2008rq,Cai:2021uup,Cai:2018dig,Odintsov:2021kup,Benetti:2021uea,Lin:2021vwc,Zhang:2021vak,Odintsov:2021urx,Pritchard:2004qp,Zhang:2005nv,Baskaran:2006qs}
and references therein. Notable is also the work of S. Weinberg in
the field, see for example Refs.
\cite{Weinberg:2003ur,Flauger:2007es}, especially notable is  the
effect of neutrino on the primordial gravitational waves, which
was first discussed in \cite{Weinberg:2003ur}. The primordial
gravitational waves form a stochastic background and on this
background, important information during and after the
inflationary era is imprinted. The stochastic primordial
gravitational wave background is a unique tool that will probe
directly the inflationary and short-post-inflationary era, since
these gravitational waves are superadiabatic amplifications of the
gravitational field's zero-point fluctuations. More importantly,
the evolution of the primordial gravitational waves is described
by linear differential equations since the coupling of the
gravitational waves with matter is tiny, contrary to the CMB
modes, which obey non-linear evolution equations for wavelengths
larger than $10$$\,$Mpc. There are three vital effects on the
primordial gravitational waves spectrum, first the effects of the
first horizon crossing during inflation, secondly the
post-inflationary second horizon crossing, where the primordial
tensor modes become subhorizon modes again, and thirdly,
post-inflationary effects on the spectrum, caused by several
sources, like matter content of the Universe, supersymmetry
breaking or even modified gravity. All these effects are encoded
on the stochastic background of primordial gravitational waves and
will certainly offer insights on the physics that stretch back
from the electroweak phase transition to the inflationary era.

If the outcome of future interferometric experiments is the
discovery of a primordial gravitational wave stochastic
background, modified gravity seems to be a compelling description
of the primordial era of our Universe, and specifically of the
inflationary and short post-inflationary ones. This is due to the
fact that standard descriptions of the inflationary era, like
scalar field theories, fail to produce an observable energy
spectrum of primordial gravitational waves
\cite{Breitbach:2018ddu}, unless tachyons are used. In the context
of modified gravity on the other hand, blue tilted inflation is
predicted, and a blue tilt can even explain recent observations on
pulsar timing arrays, see for example
\cite{Kuroyanagi:2020sfw,Vagnozzi:2020gtf}. To be specific, a blue
tilted tensor spectral index or an abnormal reheating era can
produce an observable gravitational wave spectrum. In view of the
above, in this paper we shall thoroughly study how to calculate
the modified gravity effects on the energy spectrum of primordial
gravitational waves. We shall use a WKB method firstly introduced
in Ref. \cite{Nishizawa:2017nef}, which offers the possibility to
quantify the overall effect of modified gravity on an integral of
a single parameter $a_M$. After reviewing how to calculate the
energy spectrum of the primordial gravitational waves, including
the modified gravity effects, we shall calculate the parameter
$a_M$ for several modified gravity theories of interest. Our aim
is to offer all the information needed for the calculation of
modified gravity effects for redshifts stretching from present
time up to the end of the inflationary era.

This article is organized as follows: In section II we present all
the formalism necessary for the extraction of the energy spectrum
of the primordial gravitational waves. We review standard features
of primordial gravitational waves and we extract the differential
equation which governs the evolution of the primordial
gravitational waves. We also show explicitly how the effects of
modified gravity are encoded on a single parameter and we discuss
how to calculate the overall effect of modified gravity on the
primordial gravitational waves. In section III, we calculate and
present formulas for the parameter $a_M$ which quantifies the
effect of modified gravity on primordial gravitational waves. We
shall calculate it for several modified gravities of
phenomenological interest, and specifically for $f(R,\phi)$
gravity, for Chern-Simons-corrected $f(R,\phi)$, for
Einstein-Gauss-Bonnet-corrected $f(R,\phi)$ gravity and for higher
derivative Einstein-Gauss-Bonnet-corrected $f(R,\phi)$ gravity.
Finally, the conclusions of this work are presented at the end of
the paper.

\section{The Spectrum of Primordial Gravitational Waves in General Relativity and Modified Gravity Effects}

In this section we shall review the general features of primordial
gravitational waves in the context of general relativity (GR) and
we shall also quantify the way that modified gravity affects the
spectrum. The analysis shall be based on Refs.
\cite{Boyle:2005se,Nishizawa:2017nef,Arai:2017hxj,Nunes:2018zot,Liu:2015psa,Zhao:2013bba,Odintsov:2021kup}
and references therein, and more details can be found in
\cite{Boyle:2005se}.

The primordial tensor perturbations are basically perturbations of
a flat Friedmann-Robertson-Walker (FRW) metric,
\begin{equation}
\label{metric} \centering {\rm d}s^2=-{\rm
d}t^2+a(t)^2\sum_{i=1}^{3}{({\rm d} x^{i})^2}\, ,
\end{equation}
where $a(t)$ is the scale factor as usual. Using the conformal
time $\tau$, the perturbed FRW metric is,
\begin{equation}
  {\rm d}s^{2}=a^{2}[-{\rm d}\tau^{2}+(\delta_{ij}+h_{ij})
  {\rm d}x^{i}{\rm d}x^{j}],
\end{equation}
with $x^{i}$ being the comoving spatial coordinates, and $h_{ij}$
denotes the gauge-invariant metric tensor perturbation, which is
symmetric $h_{ij}\!=\!h_{ji}$, traceless $h_{ii}\!=\!0$ and
transverse $\partial^j h_{ij}\!=\!0$ conditions. The reason for
traceless, transverse and symmetric is that every tensor mode
describing a  gravitational wave should actually have these
features. The second order Lagrangian corresponding to the tensor
perturbation $h_{ij}(\tau,{\bf x})$ is,
\begin{equation}
  \label{tensor_action}
  S=\int d\tau d{\bf x}\sqrt{-g}\left[
    \frac{-g^{\mu\nu}}{64\pi G}\partial_{\mu}h_{ij}
    \partial_{\nu}h_{ij}+\frac{1}{2}\Pi_{ij}h_{ij}\right],
\end{equation}
where the tensor part of the anisotropic stress $\Pi_{\mu \nu}$
is,
\begin{equation}
  \Pi^{i}_{j}=T^{i}_{j}-p\delta^{i}_{j}
\end{equation}
and satisfies $\Pi_{ii}=0$, $\partial^{i}\Pi_{ij}=0$, while it
acts as an external source in the action (\ref{tensor_action}).
Upon varying the action (\ref{tensor_action}) with respect to
$h_{ij}$, we obtain,
\begin{equation}
  \label{h_eq}
  h_{ij}''+2\frac{a'(\tau)}{a(\tau)}h_{ij}'-{\bf \nabla}^{2}h_{ij}
  =16\pi G a^{2}(\tau)\Pi_{ij}(\tau,{\bf x}),
\end{equation}
with the ``prime'' denoting differentiation with respect to the
conformal time. Upon Fourier transforming Eq. (\ref{h_eq}), we
get,
\begin{subequations}
  \label{fourier_expand}
  \begin{eqnarray}
    h_{ij}^{}(\tau,{\bf x})\!\!&=\!&\!\!\sum_{r}
    \sqrt{16\pi G}\!\!\int\!\!\!
    \frac{d{\bf k}}{(2\pi)^{3/2}}\epsilon_{ij}^{r}({\bf k})
    h_{{\bf k}}^{r}(\tau){\rm e}^{i{\bf k}{\bf x}},\qquad\quad \\
    \Pi_{ij}^{}(\tau,{\bf x})\!\!&=\!&\!\!\sum_{r}
    \sqrt{16\pi G}\!\!\int\!\!\!
    \frac{d{\bf k}}{(2\pi)^{3/2}}\epsilon_{ij}^{r}({\bf k})
    \Pi_{{\bf k}}^{r}(\tau){\rm e}^{i{\bf k}{\bf x}},\qquad\quad
  \end{eqnarray}
\end{subequations}
with $r=$(``$+$'' or ``$\times$'') denoting the polarization of
the tensor perturbation, and the polarization tensors satisfy
[$\epsilon_{ij}^{r} ({\bf
  k})=\epsilon_{ji}^{r}({\bf k})$], and also $\epsilon_{ii}^{r}
({\bf k})=0$, and $k_{i}\epsilon_{ij}^{r}({\bf k})=0$. Eq.
(\ref{fourier_expand}) in conjunction with Eq.
(\ref{tensor_action}) yields,
\begin{equation}
  \label{tensor_action_fourier}
  S\!=\!\!\sum_{r}\!\!\int\!\!d\tau d{\bf k}\frac{a^{2}}{2}\;\!\!
  \Big[h_{{\bf k}}^{r}{}'h_{\!\textrm{-}{\bf k}}^{r}\!{}'\!
  -\!k^{2}h_{{\bf k}}^{r}h_{\!\textrm{-}{\bf k}}^{r}\!
  +\!32\pi G a^{2}\Pi_{{\bf k}}^{r}h_{\!\textrm{-}{\bf k}}^{r}
  \Big]\, ,
\end{equation}
which is the action for the Fourier transformed gravitational
tensor perturbations. The resulting theory can easily be
quantized, with $h_{{\bf k}}^{r}$ playing the role of the
canonical variable, and the corresponding conjugate momentum is,
\begin{equation}
  \label{def_pi}
  \pi_{{\bf k}}^{r}(\tau)=a^{2}(\tau)h_{\!\textrm{-}{\bf
  k}}^{r}{}'(\tau)\, ,
\end{equation}
hence the theory is quantized if the following equal time
commutation relations hold true,
\begin{subequations}
  \label{h_pi_commutators}
  \begin{eqnarray}
    \left[\hat{h}_{{\bf k}}^{r}(\tau),
      \hat{\pi}_{{\bf k}'}^{s}(\tau)\right]
    &=&i\delta^{rs}\delta^{(3)}({\bf k}-{\bf k}'), \\
    \left[\hat{h}_{{\bf k}}^{r}(\tau),
      \hat{h}_{{\bf k}'}^{s}(\tau)\right]
    &=&\left[\hat{\pi}_{{\bf k}}^{r}(\tau),
      \hat{\pi}_{{\bf k}'}^{s}(\tau)\right]=0.
  \end{eqnarray}
\end{subequations}
The Fourier components of $\hat{h}_{ij}(\tau,{\bf x})$ satisfy the
relation $\hat{h}_{{\bf k}}^{r}=\hat{h}_{\!\textrm{-}{\bf
k}}^{r\dag}$, since $\hat{h}_{ij}(\tau,{\bf x})$ is a Hermitian
operator. Therefore we have,
\begin{equation}
  \label{h_from_a}
  \hat{h}_{{\bf k}}^{r}(\tau)=h_{k}^{}(\tau)\hat{a}_{{\bf k}}^{r}
  +h_{k}^{\ast}(\tau)\hat{a}_{\!\textrm{-}{\bf k}}^{r\dag},
\end{equation}
where $\hat{a}_{{\bf k}}^{r\dag}$ and $\hat{a}_{{\bf k}}^{r}$
being the creation and annihilation operators respectively, which
satisfy,
\begin{subequations}
  \label{a_commutators}
  \begin{eqnarray}
    \Big[\hat{a}_{{\bf k}}^{r},\hat{a}_{{\bf k}'}^{s\dag}\Big]
    &=&\delta^{rs}\delta^{(3)}({\bf k}-{\bf k}'), \\
    \Big[\hat{a}_{{\bf k}}^{r},\hat{a}_{{\bf k}'}^{s}\Big]
    &=&\Big[\hat{a}_{{\bf k}}^{r\dag},\hat{a}_{{\bf
    k}'}^{s\dag}\Big]=0\, .
  \end{eqnarray}
\end{subequations}
More importantly, each mode $h_{k}(\tau)$ satisfies the equation,
\begin{equation}
  \label{h_eq_ft}
  h_{k}''+2\frac{a'(\tau)}{a(\tau)}h_{k}'+k^{2}h_{k}^{}=
    16\pi G a^{2}(\tau)\Pi_{k}^{}(\tau).
\end{equation}
The modes $h_{k}^{}(\tau)$ depend both on the conformal time and
on the wavenumber  $k=|{\bf k}|$, but do not depend on the
polarization and on the direction, as it is apparent from Eq.
(\ref{h_from_a}). The Wronskian normalization condition,
\begin{equation}
  \label{Wronskian}
  h_{k}^{}(\tau)h_{k}^{\ast}{}'(\tau)-h_{k}^{\ast}(\tau)h_{k}'(\tau)
  =\frac{i}{a^{2}(\tau)}\, ,
\end{equation}
makes compatible the commutation relations
(\ref{h_pi_commutators}) and (\ref{a_commutators}), and also the
initial condition for the modes is,
\begin{equation}
  \label{h_bc}
  h_{k}^{}(\tau)\to\frac{{\rm exp}(-ik\tau)}{a(\tau)\sqrt{2k}}
  \qquad({\rm as}\;\;\tau\to-\infty),
\end{equation}
which corresponds to the Bunch-Davies vacuum and describes modes
which are initially at subhorizon scales and satisfy Eq.
(\ref{Wronskian}).

Let us proceed to the spectrum of the primordial gravitational
waves $\Omega_{gw}^{}(k,\tau)$, and an inherent quantity to the
energy spectrum, the tensor power spectrum
$\Delta_{h}^{2}(k,\tau)$. The latter can be obtained by
considering,
\begin{equation}
  \langle0|\hat{h}_{ij}^{}(\tau,{\bf x})\hat{h}_{ij}^{}
  (\tau,{\bf x})|0\rangle\!=\!\!\!\int_{0}^{\infty}\!\!\!\!\!64\pi G
  \frac{k^{3}}{2\pi^{2}}\!\left|h_{k}^{}(\tau)\right|^{2}
  \!\frac{dk}{k},
\end{equation}
and thus, the inflationary tensor power spectrum is obtained,
\begin{equation}
  \label{def_tensor_power}
  \Delta_{h}^{2}(k,\tau)\equiv\frac{d\langle0|\hat{h}_{ij}^{2}
    |0\rangle}{d\,{\rm ln}\,k}=64\pi G\frac{k^{3}}{2\pi^{2}}
  \left|h_{k}^{}(\tau)\right|^{2}.
\end{equation}
The energy spectrum of the primordial gravitational waves
evaluated at present day $\Omega_{gw}^{}(k,\tau)$ is equal to,
\begin{equation}
  \label{def_Omega_gw}
  \Omega_{gw}^{}(k,\tau)\equiv\frac{1}{\rho_{crit}^{}(\tau)}
  \frac{d\langle 0|\hat{\rho}_{gw}^{}(\tau)|0\rangle}{d\,{\rm
  ln}\,k}\, .
\end{equation}
By definition, the energy spectrum of the primordial gravitational
waves is the energy density of the gravitational waves, evaluated
per logarithmic wavenumber interval. For the evaluation of
$\Omega_{gw}^{}(k,\tau)$, we can treat the tensor perturbation
$h_{ij}$ as a quantum field in an unperturbed FRW geometric
background, the stress-energy tensor of which is,
\begin{equation}
  \label{T_alpha_beta}
  T_{\alpha\beta}=-2\frac{\delta L}{\delta\bar{g}^{\alpha\beta}}
  +\bar{g}_{\alpha\beta}L,
\end{equation}
as it is obtained by the action (\ref{tensor_action}), with $L$
denoting the Lagrangian function in Eq. (\ref{tensor_action}).
Since the future laser interferometers will seek for primordial
gravitational waves in high frequencies compared to the CMB ones,
we can omit the anisotropic stress couplings, and thus the
gravitational wave energy density is,
\begin{equation}
  \label{rho_gw}
  \rho_{gw}^{}=-T_{0}^{0}=\frac{1}{64\pi G}
  \frac{(h_{ij}')^{2}+(\vec{{\bf \nabla}}h_{ij})^{2}}{a^{2}},
\end{equation}
and the corresponding vacuum expectation value of it is,
\begin{equation}
  \label{rho_gw_expect}
  \langle0|\rho_{gw}^{}|0\rangle=\int_{0}^{\infty}\frac{k^{3}}
  {2\pi^{2}}\frac{\left|h_{k}'\right|^{2}
    +k^{2}\left|h_{k}^{}\right|^{2}}{a^{2}}\frac{dk}{k},
\end{equation}
hence the gravitational wave energy spectrum reads,
\begin{equation}
  \Omega_{gw}^{}(k,\tau)=\frac{8\pi G}{3H^{2}(\tau)}
  \frac{k^{3}}{2\pi^{2}}
  \frac{\left|h_{k}'(\tau)\right|^{2}
    +k^{2}\left|h_{k}^{}(\tau)\right|^{2}}{a^{2}(\tau)}.
\end{equation}
Moreover, using $|h_{k}'(\tau)|^{2}=k^{2}|h_{k}^{}(\tau)|^{2}$ the
gravitational wave energy spectrum evaluated at present day can be
rewritten as follows,
\begin{equation}
  \label{spec_relations}
  \Omega_{gw}(k,\tau)=\frac{1}{12}\frac{k^{2}\Delta_{h}^{2}(k,\tau)}
  {H_0^{2}(\tau)}\, ,
\end{equation}
where $H_0$ is the Hubble rate at present day, and we also assumed
that the present day scale factor is equal to unity, in order for
comoving quantities (frequencies and wavelengths) to be identical
with physical quantities. The interest in primordial gravitational
wave searches is for large frequency modes that became subhorizon
during the dark era of reheating and radiation domination era,
thus for modes which entered the Hubble horizon first after the
end of the inflationary era. Let us now be more concrete on the
calculation of the energy spectrum of the primordial gravity
waves, and first we consider the effects on it, caused by horizon
re-entry of a mode $k$, in which case
\cite{Boyle:2005se,Nishizawa:2017nef,Arai:2017hxj,Nunes:2018zot,Liu:2015psa,Zhao:2013bba},
\begin{equation}
    h_k^{\lambda}(\tau)=h_k^{\lambda ({\rm p})}
    \left( \frac{3j_1(k\tau)}{k\tau}\right),
\end{equation}
where $j_{\ell}$ denotes the $\ell$-th spherical Bessel function.
The Fourier transformation of the primordial tensor perturbation
during a power-law cosmological evolution $a(t)\propto t^p$ is,
\begin{equation}
    h_k(\tau) \propto
    a(t)^{\frac{1-3p}{2p}}J_{\frac{3p-1}{2(1-p)}}( k\tau ),
\end{equation}
with $J_n(x)$ being the Bessel function. Moreover, taking into
account the damping caused by the relativistic degrees of freedom
in the early Universe which do not remain constant, the following
factor for $h_k(\tau)$ is obtained \cite{Watanabe:2006qe},
\begin{equation}
    \left ( \frac{g_*(T_{\rm in})}{g_{*0}} \right )
    \left ( \frac{g_{*s0}}{g_{*s}(T_{\rm in})} \right )^{4/3},
\end{equation}
where the scale factor evolves as $a(t) \propto T^{-1}$ assuming
adiabatic evolution which follows by the fact that the entropy of
the Universe $S$ is constant,
\begin{equation}\label{entropy}
S=\frac{2\pi^2}{4s}g_{*s}\left( a T \right)^3=\mathrm{const}\, .
\end{equation}
Above, $T_{\rm in}$ denotes the temperature of the Universe at
horizon re-entry,
\begin{equation}
    T_{\rm in}\simeq 5.8\times 10^6~{\rm GeV}
    \left ( \frac{g_{*s}(T_{\rm in})}{106.75} \right )^{-1/6}
    \left ( \frac{k}{10^{14}~{\rm Mpc^{-1}}} \right ).
\end{equation}
Also it is vital to take into account the damping factor caused by
the current acceleration of the Universe $\sim
(\Omega_m/\Omega_\Lambda)^2$ \cite{Zhang:2005nw,Boyle:2005se}. The
full expression for the present day measured primordial gravity
waves energy spectrum per log frequency, includes all the
aforementioned damping effects and in addition and more
importantly, it contains the transfer functions which are
calculated numerically by integrating the evolution differential
equation of the Fourier transformed tensor perturbation. The full
expression for the energy density of the primordial gravitational
waves is,
\begin{equation}
    \Omega_{\rm gw}(f)= \frac{k^2}{12H_0^2}\Delta_h^2(k),
    \label{GWspec}
\end{equation}
with the detailed form of $\Delta_h^2(k)$ being
\cite{Boyle:2005se,Nishizawa:2017nef,Arai:2017hxj,Nunes:2018zot,Liu:2015psa,Zhao:2013bba},
\begin{equation}
\Delta_h^2(k)=\Delta_h^{({\rm p})}(k)^{2} \left (
\frac{\Omega_m}{\Omega_\Lambda} \right )^2 \left (
\frac{g_*(T_{\rm in})}{g_{*0}} \right ) \left (
\frac{g_{*s0}}{g_{*s}(T_{\rm in})} \right )^{4/3} \left
(\overline{ \frac{3j_1(k\tau_0)}{k\tau_0} } \right )^2 T_1^2\left
( x_{\rm eq} \right ) T_2^2\left ( x_R \right )\, ,
\label{mainfunctionforgravityenergyspectrum}
\end{equation}
where $g_*(T_{\mathrm{in}}(k))$ in Eq.
(\ref{mainfunctionforgravityenergyspectrum}) is
\cite{Kuroyanagi:2014nba},
\begin{align}\label{gstartin}
& g_*(T_{\mathrm{in}}(k))=g_{*0}\left(\frac{A+\tanh \left[-2.5
\log_{10}\left(\frac{k/2\pi}{2.5\times 10^{-12}\mathrm{Hz}}
\right) \right]}{A+1} \right) \left(\frac{B+\tanh \left[-2
\log_{10}\left(\frac{k/2\pi}{6\times 10^{-19}\mathrm{Hz}} \right)
\right]}{B+1} \right)\, ,
\end{align}
with the parameters $A$ and $B$ being equal to,
\begin{equation}\label{alphacap}
A=\frac{-1-10.75/g_{*0}}{-1+10.75g_{*0}}\, ,
\end{equation}
\begin{equation}\label{betacap}
B=\frac{-1-g_{max}/10.75}{-1+g_{max}/10.75}\, ,
\end{equation}
with $g_{max}=106.75$ and $g_{*0}=3.36$. Also by replacing
$g_{*0}=3.36$ with $g_{*s}=3.91$, we can calculate
$g_{*0}(T_{\mathrm{in}}(k))$, using the same formulas, namely,
Eqs. (\ref{gstartin}), (\ref{alphacap}) and (\ref{betacap}). Note
that the ``bar'' above the Bessel function in Eq.
(\ref{mainfunctionforgravityenergyspectrum}) indicates averaging
over many integration periods. Also, the term $\Delta_h^{({\rm
p})}(k)^{2}$ in Eq. (\ref{mainfunctionforgravityenergyspectrum})
denotes the inflationary era's primordial tensor spectrum, the
analytic form of which is
\cite{Boyle:2005se,Nishizawa:2017nef,Arai:2017hxj,Nunes:2018zot,Liu:2015psa,Zhao:2013bba},
\begin{equation}
\Delta_h^{({\rm
p})}(k)^{2}=\mathcal{A}_T(k_{ref})\left(\frac{k}{k_{ref}}
\right)^{n_T}\, , \label{primordialtensorpowerspectrum}
\end{equation}
where the scale $k_{ref}=0.002$$\,$Mpc$^{-1}$ is the CMB pivot
scale. Also $\mathcal{A}_T(k_{ref})$ denotes the amplitude of the
primordial tensor perturbations, $n_T$ denotes the tensor spectral
index. Replacing,
\begin{equation}\label{amplitudeoftensorperturbations}
\mathcal{A}_T(k_{ref})=r\mathcal{P}_{\zeta}(k_{ref})\, ,
\end{equation}
where $\mathcal{P}_{\zeta}(k_{ref})$ is the amplitude of the
primordial scalar perturbations, we have finally,
\begin{equation}\label{primordialtensorspectrum}
\Delta_h^{({\rm
p})}(k)^{2}=r\mathcal{P}_{\zeta}(k_{ref})\left(\frac{k}{k_{ref}}
\right)^{n_T}\, .
\end{equation}
Regarding the transfer functions $T_1(x_{\rm eq})$ and $T_2(x_R)$,
the analytic form of the first one is
\cite{Boyle:2005se,Nishizawa:2017nef,Arai:2017hxj,Nunes:2018zot,Liu:2015psa,Zhao:2013bba},
\begin{equation}
    T_1^2(x_{\rm eq})=
    \left [1+1.57x_{\rm eq} + 3.42x_{\rm eq}^2 \right ], \label{T1}
\end{equation}
and it basically characterizes the modes that re-entered the
Hubble horizon approximately during the matter-radiation equality
with cosmic time instance $t=t_{\rm eq}$. About the notation of
several parameters in Eq. (\ref{T1}), $x_{\rm eq}=k/k_{\rm eq}$
and $k_{\rm eq}\equiv a(t_{\rm eq})H(t_{\rm eq}) = 7.1\times
10^{-2} \Omega_m h^2$ Mpc$^{-1}$. Now regarding the transfer
function $T_2(x_R)$ appearing in Eq.
(\ref{mainfunctionforgravityenergyspectrum}), it is related with
modes that entered the Hubble horizon during the reheating era,
with the mode $k$ being $k>k_R$. Its analytic form is,
\begin{equation}\label{transfer2}
 T_2^2\left ( x_R \right )=\left(1-0.22x^{1.5}+0.65x^2
 \right)^{-1}\, ,
\end{equation}
where $x_R=\frac{k}{k_R}$, and also the $k_R$ wavenumber is equal
to,
\begin{equation}
    k_R\simeq 1.7\times 10^{13}~{\rm Mpc^{-1}}
    \left ( \frac{g_{*s}(T_R)}{106.75} \right )^{1/6}
    \left ( \frac{T_R}{10^6~{\rm GeV}} \right )\, ,  \label{k_R}
\end{equation}
with $T_R$ being the reheating temperature. Also the reheating
frequency is,
\begin{equation}
    f_R\simeq 0.026~{\rm Hz}
    \left ( \frac{g_{*s}(T_R)}{106.75} \right )^{1/6}
    \left ( \frac{T_R}{10^6~{\rm GeV}} \right ).  \label{f_R}
\end{equation}
Having presented the GR primordial gravitational wave energy
density, in the next section we shall consider the effects caused
by modified gravity of various forms on the GR waveform. Thus we
shall quantify the effects of modified gravity on the GR waveform
in an explicit way.

\subsection{The Modified Gravity Effect on the Energy Spectrum of the Primordial Gravitation Waves: A WKB Approach}

\begin{figure}
\centering
\includegraphics[width=22pc]{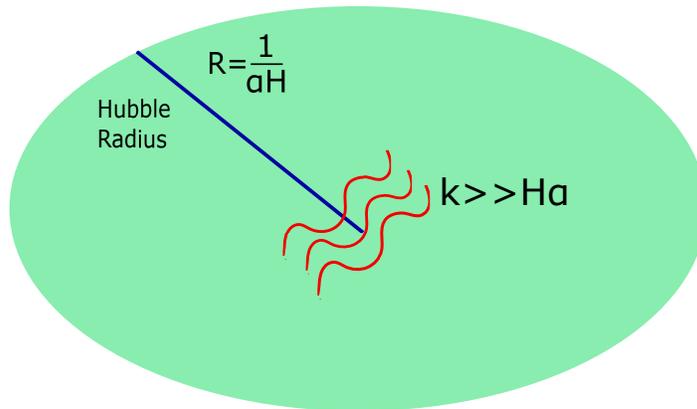}
\caption{Post-inflationary subhorizon modes for which the WKB
solution of the evolution equation
(\ref{mainevolutiondiffeqnfrgravityconftime}) is justified. These
subhorizon modes, became subhorizon immediately after inflation,
so during the early stages of the dark ages, the reheating and the
subsequent radiation domination eras. These early subhorizon modes
will be probed by the future interferometer gravitational wave
experiments.}\label{plot1}
\end{figure}

In this section we shall quantify the effect of an arbitrary
modified gravity on the GR waveform of primordial gravitational
waves. Let us recall for convenience at this point the
differential equation that is obeyed by the Fourier transformation
of the tensor perturbation $h_{i j}$,
\begin{equation}\label{mainevolutiondiffeqnfrgravity}
\ddot{h}(k)+\left(3+a_M \right)H\dot{h}(k)+\frac{k^2}{a^2}h(k)=0\,
,
\end{equation}
with $\alpha_M$ being defined as follows,
\begin{equation}\label{amfrgravity}
a_M=\frac{\dot{Q}_t}{Q_tH}\, ,
\end{equation}
where $Q_t$ is unique for every distinct modified gravity. The
overall effect of modified gravity is encoded on the parameter
$a_M$, and the functional form of this parameter is different for
distinct modified gravities. In addition, the evolution
differential equation (\ref{mainevolutiondiffeqnfrgravity})
characterizes all the distinct polarizations of the gravitational
waves. In order to extract in a consistent way the overall
modified gravity effect, we shall use Nishizawa's approach
\cite{Nishizawa:2017nef,Arai:2017hxj}, which is basically a WKB
approach. Expressed in terms of the conformal time, the
differential equation (\ref{mainevolutiondiffeqnfrgravity})
becomes,
\begin{equation}\label{mainevolutiondiffeqnfrgravityconftime}
h''(k)+\left(2+a_M \right)\mathcal{H} h'(k)+k^2h(k)=0\, ,
\end{equation}
with the ``prime'' in the above equation indicating
differentiation with respect to the conformal time $\tau$, and
also we defined $\mathcal{H}=\frac{a'}{a}$. We shall extract the
WKB solution by taking into account only subhorizon modes
satisfying Eq. (\ref{mainevolutiondiffeqnfrgravityconftime}) (see
Fig. \ref{plot1}). This is highly justified for primordial
gravitational waves studies, since the high frequency
interferometers will probe modes which became subhorizon modes
immediately after the inflationary era, so during the early
reheating era.
Considering a solution of the form
$h_{ij}=\mathcal{A}e^{i\mathcal{B}}h_{ij}^{GR}$, for theories in
which the speed of the gravitational wave is equal to unity in
natural units, the WKB solution for subhorizon modes is
\cite{Nishizawa:2017nef,Arai:2017hxj},
(\ref{mainevolutiondiffeqnfrgravityconftime}) is of the form,
\begin{equation}\label{mainsolutionwkb}
h=e^{-\mathcal{D}}h_{GR}\, ,
\end{equation}
where $h_{i j}=h e_{i j}$, with $h_{GR}$ denoting the GR waveform
which is the solution to the differential equation
(\ref{mainevolutiondiffeqnfrgravityconftime}) by taking $a_M=0$.
More importantly, the parameter $\mathcal{D}$ is equal to,
\begin{equation}\label{dform}
\mathcal{D}=\frac{1}{2}\int^{\tau}a_M\mathcal{H}{\rm
d}\tau_1=\frac{1}{2}\int_0^z\frac{a_M}{1+z'}{\rm d z'}\, ,
\end{equation}
and mainly quantifies the direct effect of modified gravity on the
GR waveform of primordial gravitational waves. Now if one wishes
to calculate the energy spectrum of the primordial gravitational
waves, it is vital to calculate the damping/amplification factor
of Eq. (\ref{dform}) starting from redshift $z=0$ which
corresponds to present time up to redshifts corresponding deeply
in the reheating era. The latter redshifts correspond to
primordial modes that became subhorizon modes immediately after
the inflationary era, so basically we are interested in extreme
subhorizon modes at present day, with significantly small
wavelength compared to the CMB scale modes. These subhorizon modes
will be probed in fifteen years from now, from LISA, BBO, DECIGO
and other gravitational waves experiments. Thus, by also taking
intro account the effects of modified gravity, the energy spectrum
of the primordial gravitational waves at present day is,
\begin{align}
\label{GWspecfR}
    &\Omega_{\rm gw}(f)=e^{-2\mathcal{D}}\times \frac{k^2}{12H_0^2}r\mathcal{P}_{\zeta}(k_{ref})\left(\frac{k}{k_{ref}}
\right)^{n_T} \left ( \frac{\Omega_m}{\Omega_\Lambda} \right )^2
    \left ( \frac{g_*(T_{\rm in})}{g_{*0}} \right )
    \left ( \frac{g_{*s0}}{g_{*s}(T_{\rm in})} \right )^{4/3} \nonumber  \left (\overline{ \frac{3j_1(k\tau_0)}{k\tau_0} } \right )^2
    T_1^2\left ( x_{\rm eq} \right )
    T_2^2\left ( x_R \right )\, .
\end{align}
Depending on the specific form of the modified gravity, the
parameter $\mathcal{D}$ might be positive or negative. Therefore,
the GR energy spectrum might be damped or amplified due to the
overall modified gravity effects. In the next subsections we shall
calculate in detail all the possible forms of the parameter $a_M$
appearing in Eqs. (\ref{amfrgravity}) and
(\ref{mainevolutiondiffeqnfrgravityconftime}) corresponding to
various modified gravities of interest. We shall give the
expressions of $a_M$ both with respect to the cosmic time and with
respect to the redshift, for the calculational convenience of the
reader. We shall also present the results in a table also for
reading convenience of the reader.

\section{Primordial Gravity Waves in Modified Gravity in its Various Forms}

Let us first consider the calculation of the parameter $a_M$
appearing in Eq. (\ref{amfrgravity}) for the case of pure
$f(R,\phi)$ gravity, in which case the gravitational action is,
\begin{equation}\label{action1}
\mathcal{S}=\int
\mathrm{d}^4x\sqrt{-g}\Big{(}\frac{f(R,\phi)}{2}-\frac{\omega(\phi)}{2}\partial^{\mu}\phi\partial_{\mu}\phi-V(\phi)\Big{)}\,
.
\end{equation}
For this form of modified gravity, the parameter $Q_t$ is equal to
$Q_t=\frac{1}{\kappa^2}\frac{\partial f(R,\phi)}{\partial R}$,
where $\kappa=\frac{1}{M_p}$, with $M_p$ being the reduced Planck
mass. Hence, for pure $f(R,\phi)$ gravity, the parameter $a_M$
reads,
\begin{equation}\label{amfrphi}
a_M=\frac{\frac{\partial^2f}{\partial R \partial
\phi}\dot{\phi}+\frac{\partial^2f}{\partial
R^2}\dot{R}}{\frac{\partial f}{\partial R}H}\, .
\end{equation}
We can express the above formula in terms of the redshift in order
to have an expression ready for the integral in Eq. (\ref{dform})
in terms of the redshift. Using the following formula,
\begin{equation}\label{formulaforredhsift}
\frac{\mathrm{d}}{\mathrm{d}t}=-H(1+z)\frac{\mathrm{d}}{\mathrm{d}z}\,
,
\end{equation}
the parameter $a_M$ expressed in terms of the redshift reads,
\begin{equation}\label{amrfrphiredshift}
a_M=\frac{-\frac{\partial^2f}{\partial R \partial
\phi}H(z)(1+z)\frac{\mathrm{d}\phi}{\mathrm{d}z}-\frac{\partial^2f}{\partial
R^2}H(z)(1+z)\frac{\mathrm{d}R}{\mathrm{d}z}}{\frac{\partial
f}{\partial R}H}\, .
\end{equation}
Let us consider the case of a Chern-Simons corrected $f(R,\phi)$
gravity, in which case the gravitational action reads,
\begin{equation}\label{action2}
\mathcal{S}=\int
\mathrm{d}^4x\sqrt{-g}\Big{(}\frac{f(R,\phi)}{2}-\frac{\omega(\phi)}{2}\partial^{\mu}\phi\partial_{\mu}\phi-V(\phi)+\frac{1}{8}\nu
(\phi)R\tilde{R}\Big{)}\, ,
\end{equation}
where $R\tilde{R}=\epsilon^{abcd}R_{ab}^{ef}R_{cdef}$ and
$\epsilon^{abcd}$ stands for the totally antisymmetric Levi-Civita
tensor. In the literature, terms of the form $\nu
(\phi)\tilde{R}R$ are known as Chern-Simons terms. We need to note
though that the term $\nu (\phi)\tilde{R}R$ is formally the
Chern-Pontryagin density, which connected to an actual three
dimensional Chern-Simons term via the exterior derivative $\nu
(\phi)\tilde{R}R=d(Chern-Simons)$. Notably the Chern-Pontryagin
density is the analogue of the quantity $ ^*F_{\mu \nu}F^{\mu
\nu}$ which is constructed by the curvature $F_{\mu \nu}$ on a
principal bundle with connection $A_{\mu}$, but abusively in the
literature it is called Chern-Simons term, due to the analogy we
pointed out. In the Chern-Simons corrected $f(R,\phi)$ gravity,
the term $Q_t$ is equal to \cite{Hwang:2005hb},
\begin{equation}\label{qttermchernsimons}
Q_t=\frac{1}{\kappa^2}\frac{\partial f}{\partial
R}+\frac{2\lambda_{\ell}\dot{\nu}k}{a}\, ,
\end{equation}
where $\lambda_{\ell}$ denotes the polarization of the
gravitational wave and takes the values $\lambda_{R}=1$ for right
handed gravity waves, and $\lambda_{L}=-1$ for left handed gravity
waves, while $k$ is the wavenumber of the tensor mode. Thus for
the Chern-Simons corrected $f(R,\phi)$ gravity, the $a_M$ term in
terms of the cosmic time reads,
\begin{equation}\label{amfrphichernsimons}
a_M=\frac{\frac{1}{\kappa^2}\frac{\partial^2f}{\partial R \partial
\phi}\dot{\phi}+\frac{1}{\kappa^2}\frac{\partial^2f}{\partial
R^2}\dot{R}+\frac{2\lambda_{\ell}\ddot{\nu}k}{a}-\frac{2\lambda_{\ell}\dot{\nu}k\,H}{a}}{\left(\frac{1}{\kappa^2}\frac{\partial
f}{\partial R}+\frac{2\lambda_{\ell}\dot{\nu}k}{a}\right)H}\, ,
\end{equation}
where the last term numerator is found by differentiating $Q_t$ in
Eq. (\ref{qttermchernsimons}). Expressing $a_M$ in terms of the
redshift, we have,
\begin{equation}\label{amrfrphiredshiftchernsimons}
a_M=\frac{-\frac{1}{\kappa^2}\frac{\partial^2f}{\partial R
\partial \phi}H(z)(1+z)\frac{\mathrm{d}\phi}{\mathrm{d}z}-\frac{1}{\kappa^2}\frac{\partial^2f}{\partial
R^2}H(z)(1+z)\frac{\mathrm{d}R}{\mathrm{d}z}+\frac{2\lambda_{\ell}H^2(1+z)\frac{\mathrm{d}\nu}{\mathrm{d}z}k}{a}+\frac{2\lambda_{\ell}\nu_{dd}(z)k}{a}}
{\left(\frac{1}{\kappa^2}\frac{\partial f}{\partial
R}-\frac{2\lambda_{\ell}(1+z)H\frac{\mathrm{d}\nu}{\mathrm{d}z}}{a}\right)H}\,
,
\end{equation}
where $\nu_{dd}(z)$ is equal to,
\begin{equation}\label{nudd}
\nu_{dd}(z)=-(1+z)H\left(\frac{\mathrm{d}H}{\mathrm{d}z}(1+z)\frac{\mathrm{d}\nu}{\mathrm{d}z}+\frac{\mathrm{d}\nu}{\mathrm{d}z}H+H(1+z)\frac{\mathrm{d}^2\nu}{\mathrm{d}z^2}
\right)\, .
\end{equation}
Now let us consider the case of $f(R,\phi)$ gravity with an
Einstein-Gauss-Bonnet term, in which case the gravitational action
reads,
\begin{equation}\label{action3}
\mathcal{S}=\int
\mathrm{d}^4x\sqrt{-g}\Big{(}\frac{f(R,\phi)}{2}-\frac{\omega(\phi)}{2}\partial^{\mu}\phi\partial_{\mu}\phi-V(\phi)-\frac{1}{2}\xi(\phi)\mathcal{G}\Big{)}\,
.
\end{equation}
For these theories, the speed of gravitational tensor
perturbations is not equal to that of light's, but it is equal to
\cite{Hwang:2005hb},
\begin{equation}\label{gravitationalwavespeed}
c_T^2=1-\frac{4\left(\ddot{\xi}-\dot{\xi}H \right)}{\frac{\partial
f}{\partial R}-4\dot{\xi}H}\, .
\end{equation}
Since we shall consider only theories for which the gravitational
wave speed is equal to that of light's, the condition $c_T^2=1$
can be satisfied only if the Gauss-Bonnet scalar coupling function
$\xi (\phi)$ satisfies the differential equation
$\ddot{\xi}-\dot{\xi}H =0$. The inflationary phenomenology of this
class of Einstein-Gauss-Bonnet theories have been thoroughly
studied in the literature, see
\cite{Odintsov:2020sqy,Odintsov:2020zkl,Oikonomou:2020oil,Oikonomou:2021kql}.
For Einstein-Gauss-Bonnet corrected $f(R,\phi)$ theories, the
parameter $Q_t$ reads \cite{Hwang:2005hb},
\begin{equation}\label{qteinsteingaussbonnet}
Q_t=\frac{\partial f}{\partial R}-4\dot{\xi}H\, ,
\end{equation}
therefore, the parameter $a_M$ for the Einstein-Gauss-Bonnet
corrected $f(R,\phi)$ theory reads,
\begin{equation}\label{amEGB}
a_M=\frac{\frac{1}{\kappa^2}\frac{\partial^2f}{\partial R \partial
\phi}\dot{\phi}+\frac{1}{\kappa^2}\frac{\partial^2f}{\partial
R^2}\dot{R}-4\ddot{\xi}H-4\dot{\xi}\dot{H}}{\left(\frac{1}{\kappa^2}\frac{\partial
f}{\partial R}-4\dot{\xi}H \right)H}\, ,
\end{equation}
and $\ddot{\xi}$ must be replaced with $\ddot{\xi}=\dot{\xi}H$ due
to the gravitational wave speed constraint $c_T^2=1$. In terms of
the redshift, the parameter $a_M$ has the following form,
\begin{equation}\label{amrfrphiegbredshift}
a_M=\frac{-\frac{1}{\kappa^2}\frac{\partial^2f}{\partial R
\partial \phi}H(z)(1+z)\frac{\mathrm{d}\phi}{\mathrm{d}z}-\frac{1}{\kappa^2}\frac{\partial^2f}{\partial
R^2}H(z)(1+z)\frac{\mathrm{d}R}{\mathrm{d}z}+4H^2(1+z)\frac{\mathrm{d}\xi}{\mathrm{d}\phi}\frac{\mathrm{d}\phi}{\mathrm{d}z}-4H^2(1+z)^2\frac{\mathrm{d}\xi}{\mathrm{d}\phi}\frac{\mathrm{d}\phi}{\mathrm{d}z}\frac{\mathrm{d}H}{\mathrm{d}z}}
{\left(\frac{1}{\kappa^2}\frac{\partial f}{\partial
R}+4H^2(1+z)\frac{\mathrm{d}\xi}{\mathrm{d}\phi}\frac{\mathrm{d}\phi}{\mathrm{d}z}\right)H}\,
,
\end{equation}
\begin{table}[h!]
  \begin{center}
    \caption{\emph{\textbf{Forms of the Parameter $a_M$ for the Various Modified Gravities}}}
    \label{table1}
    \begin{tabular}{|r|r|}
     \hline  \textbf{Modified Gravity Type}
       & \bf{$a_M$}

      \\ \hline
        Pure $f(R,\phi)$ & $ a_M=\left(\frac{\partial^2f}{\partial R \partial
\phi}\dot{\phi}+\frac{\partial^2f}{\partial
R^2}\dot{R}\right)/\left(\frac{\partial f}{\partial R}H\right)$
      \\  \hline
 C-S  $f(R,\phi)$  & $a_M=\left(\frac{1}{\kappa^2}\frac{\partial^2f}{\partial R \partial
\phi}\dot{\phi}+\frac{1}{\kappa^2}\frac{\partial^2f}{\partial
R^2}\dot{R}+\frac{2\lambda_{\ell}\ddot{\nu}k}{a}-\frac{2\lambda_{\ell}\dot{\nu}k\,H}{a}\right)/\left(\frac{1}{\kappa^2}\frac{\partial
f}{\partial R}+\frac{2\lambda_{\ell}\dot{\nu}k}{a}\right)H$
      \\  \hline
   EGB  $f(R,\phi)$   & $a_M=\left(\frac{1}{\kappa^2}\frac{\partial^2f}{\partial R \partial
\phi}\dot{\phi}+\frac{1}{\kappa^2}\frac{\partial^2f}{\partial
R^2}\dot{R}-4\ddot{\xi}H-4\dot{\xi}\dot{H}\right)/\left(\frac{1}{\kappa^2}\frac{\partial
f}{\partial R}-4\dot{\xi}H \right)H$
      \\  \hline
     H-EGB   $f(R,\phi)$   & $a_M=\left(\frac{1}{\kappa^2}\frac{\partial^2f}{\partial R \partial
\phi}\dot{\phi}+\frac{1}{\kappa^2}\frac{\partial^2f}{\partial
R^2}\dot{R}-4c_1\ddot{\xi}H-4c_1\dot{\xi}\dot{H}+\frac{c_2}{2}\dot{\xi}\dot{\phi}^2+c_2\xi
\dot{\phi}\ddot{\phi}\right)/\left(\frac{1}{\kappa^2}\frac{\partial
f}{\partial R}-4c_1\dot{\xi}H+\frac{c_2}{2}\xi\dot{\phi}^2
\right)H$
\\
\hline
    \end{tabular}
  \end{center}
\end{table}
Finally, let us consider a generalized
Einstein-Gauss-Bonnet-corrected $f(R,\phi)$ theory with higher
order derivative couplings, in which case the gravitational action
is,
\begin{equation}\label{action4}
\mathcal{S}=\int
\mathrm{d}^4x\sqrt{-g}\Big{(}\frac{f(R,\phi)}{2}-\frac{\omega(\phi)}{2}\partial^{\mu}\phi\partial_{\mu}\phi-V(\phi)-\frac{c_1}{2}\xi(\phi)\mathcal{G}-\frac{c_2}{2}\xi(\phi)G^{\mu
\nu}\partial_{\mu}\phi\partial_{\nu}\phi\Big{)}\, ,
\end{equation}
where $G^{\mu \nu}$ is the Einstein tensor and $c_1$, $c_2$ are
dimensionless constants. For this theory, the gravitational wave
speed is again non-trivial and different from that of light's in
vacuum. Specifically the gravitational wave speed is
\cite{Hwang:2005hb},
\begin{equation}\label{gravitationalwavespeedderivative}
c_T^2=1-\frac{8c_1\left(\ddot{\xi}-\dot{\xi}H \right)+2c_2\xi
\dot{\phi}^2}{\frac{\partial f}{\partial R}-4\dot{\xi}H}\, .
\end{equation}
As in the in Einstein-Gauss-Bonnet-corrected theory we presented
previously, we are interested for theories with gravitational wave
speed equal to that of light's in vacuum, and for the case at
hand, the theory can have $c_T^2=1$ when the Gauss bonnet scalar
coupling $\xi (\phi)$ satisfies the differential equation
\begin{equation}\label{differentialequationderivative}
8c_1\left(\ddot{\xi}-\dot{\xi}H\right)+2c_2\xi \dot{\phi}^2 =0\, .
\end{equation}
With this constraint satisfied, the only parameter that needs to
be calculated for evaluating the energy spectrum of the primordial
gravitational waves spectrum is $a_M$ given in Eq.
(\ref{amfrgravity}). In the case at hand, the parameter $Q_t$ is
equal to \cite{Hwang:2005hb},
\begin{equation}\label{qteinsteingaussbonnetderivative}
Q_t=\frac{\partial f}{\partial R}-4c_1\dot{\xi}H-\frac{c_2}{2}\xi
\dot{\phi}^2\, ,
\end{equation}
therefore, the parameter $a_M$ (\ref{amfrgravity}) for the
Einstein-Gauss-Bonnet-corrected $f(R,\phi)$ theory with higher
derivative coupling terms reads,
\begin{equation}\label{amEGBderivative}
a_M=\frac{\frac{1}{\kappa^2}\frac{\partial^2f}{\partial R \partial
\phi}\dot{\phi}+\frac{1}{\kappa^2}\frac{\partial^2f}{\partial
R^2}\dot{R}-4c_1\ddot{\xi}H-4c_1\dot{\xi}\dot{H}+\frac{c_2}{2}\dot{\xi}\dot{\phi}^2+c_2\xi
\dot{\phi}\ddot{\phi} }{\left(\frac{1}{\kappa^2}\frac{\partial
f}{\partial R}-4c_1\dot{\xi}H+\frac{c_2}{2}\xi\dot{\phi}^2
\right)H}\, .
\end{equation}
The above concludes the most complicated extension of $f(R,\phi)$
modified gravity with $c_T^2=1$, for which the parameter $a_M$ of
Eq. (\ref{amfrgravity}) can be calculated. It is always a
computational challenge to calculate numerically the parameter
$\mathcal{D}$ appearing in Eq. (\ref{dform}) for redshifts
corresponding to modes which became subhorizon after the
inflationary era, during the reheating and the radiation
domination era. In Table \ref{table1} we gather all the various
forms of the parameter $a_M$ for the various forms of modified
gravity which were considered in this section.

\section{Conclusions}

In this work we studied the way that various forms of modified
gravity affect the energy spectrum of the primordial gravitational
waves. We presented the standard features of the energy spectrum
of the primordial gravitational waves in GR, and how modified
gravity affects the spectrum in a quantitative way. The critical
effect of modified gravity on the energy spectrum of the
primordial gravitational waves is quantified on a single parameter
denoted $a_M$ and its integral for redshifts extending from
present day up to redshifts corresponding to the early
post-inflationary era. We calculated the parameter $a_M$ for
several modified gravities of interest, and specifically for the
$f(R,\phi)$ gravity, for Chern-Simons-corrected $f(R,\phi)$
gravity, for Einstein-Gauss-Bonnet-corrected $f(R,\phi)$ gravity
and for higher derivative extensions of
Einstein-Gauss-Bonnet-corrected $f(R,\phi)$ gravity. The
motivation for studying several modified gravity effects on
primordial gravitational waves is based on the possible
verification of the stochastic primordial gravitational waves
background by future experiments. The actual verification will
stir things up significantly in theoretical cosmology, since
theorists must think how such a signal is generated. This is due
to the fact that standard theories of inflation, like scalar field
theory, do not produce a detectable signal of stochastic
primordial gravity wave background. Thus, a possible detection
will indicate either that some modified gravity controls the
physics of inflation and post-inflation eras, or that some
alternative reheating mechanism controls the post-inflationary
physics. However, in the context of GR, the abnormal reheating
effects could be minor and could not amplify the spectrum
significantly to be detectable, see for example  the $w$ EoS
post-inflationary parameter $C_2$ of \cite{Boyle:2005se}. Thus, it
is rather compelling to study in detail and thoroughly all the
effects of modified gravity in its various forms, on the energy
spectrum of primordial gravitational waves. With this paper we
presented a rigid overview of the method needed for calculating in
a formally correct way the overall effect of several modified
gravities of interest on the energy spectrum of the primordial
gravitational waves.

\section*{Acknowledgments}

This paper was supported by the Ministry of Education and Science
of Kazakhstan, Grant AP09261147.

\end{document}